\documentclass[journal] {IEEEtran}

\usepackage{epsfig}
\usepackage{amsmath}

\def\e{\begin{equation}}
\def\f{\end{equation}}
\def\%#1{\mbox{\boldmath $#1$}}
\def\=#1{\overline{\overline #1}}

\def\_#1{{\bf #1}}

\def\E{\epsilon}

\def\.{\cdot}

\def\##1{{\bf#1\mit}}

\def\l#1{\label{eq:#1}}
\def\r#1{(\ref{eq:#1})}
\def\am{\left(\begin{array}{c}}
\def\amm{\left(\begin{array}{cc}}
\def\a{\end{array}\right)}

\begin{document}

\title{Generalized Permeability Function and Field Energy Density in Artificial Magnetics
Using the Equivalent Circuit Method}

\author{Pekka~Ikonen,~\IEEEmembership{Student Member,~IEEE,} and
        Sergei~Tretyakov,~\IEEEmembership{Senior Member,~IEEE}
\thanks{The authors are with the Radio Laboratory / SMARAD Center of Excellence,
Helsinki University of Technology, P.O. Box 3000, FI-02015 TKK, Finland. (e-mail: pekka.ikonen@tkk.fi).}}

\maketitle

\begin{abstract}

The equivalent circuit model for artificial magnetic materials based on various arrangements of split rings is
generalized by taking into account losses in the substrate or matrix material. It is shown that a modification
is needed to the known macroscopic permeability function in order to correctly describe these materials.
Depending on the dominating loss mechanism (conductive losses in metal parts or dielectric losses in the
substrate) the permeability function has different forms. The proposed circuit model and permeability function
are experimentally validated. Furthermore, starting from the generalized circuit model we derive an explicit
expression for the electromagnetic field energy density in artificial magnetic media. This expression is valid
at low frequencies and in the vicinity of the resonance also when dispersion and losses in the material are
strong. The presently obtained results for the energy density are compared with the results obtained using
different methods.

\end{abstract}


\section{Introduction}

Artificial electromagnetic media with extraordinary properties (often called {\itshape metamaterials}) attract
increasing attention in the microwave community. One of the widely studied subclasses of metamaterials are
artificial magnetic materials operating in the microwave regime, e.g.~\cite{Kostin}--\cite{Maslovski}. Broken
loops have been used as one of the building blocks to implement double-negative (DNG) media \cite{Smith,Shelby},
in addition to this, artificial magneto-dielectric substrates are nowadays considered as one of the most
promising ways to miniaturize microstrip antennas \cite{Hansen}--\cite{Pekka}.

The extraordinary features of metamaterials call for careful
analysis when studying the fundamental electromagnetic quantities in
these materials. Recently, a lot of research has been devoted to the
definition of field energy density in DNG media
\cite{Ruppin}--\cite{Boardman}. Authors of
\cite{Ruppin,Cui,Boardman} derived the energy density expression
starting from the macroscopic media model and writing down the
equation of motion for polarization (electric charge) or
magnetization density in the media. Furthermore, complex Poynting
theorem was used to search for expressions having the mathematically
correct form to be identified as energy densities. Following the
terminology presented in \cite{Boardman} we call this method
``electrodynamic method'' (ED). Tretyakov used in
\cite{Sergei_energy} another method: Starting from the material
microstructure, an equivalent circuit representation was derived for
the unit cell constituting specific artificial dielectric and
magnetic media. Lattices of thin wires and arrays of split-ring
resonators were considered in \cite{Sergei_energy}. The stored
reactive energy, and, furthermore, the field energy density, were
calculated using the classical circuit theory. Authors of
\cite{Boardman} later called this method ``equivalent circuit
method'' (EC). Though these two methods apply to media with the same
macroscopic permeabilities and permittivities, they are
fundamentally very different, as will be clarified later in more
detail. Moreover, the final expressions for the field energy density
in artificial magnetic media given in \cite{Sergei_energy} and
\cite{Boardman} differ from each other. One of the motivations of
this work is to clarify the reasons for this difference.

Here we concentrate only on artificial magnetic media and set two
main goals for our work: 1) To understand the differences and
assumptions behind the ED and EC-methods when deriving the field
energy density expressions. We verify using a specific example
(magnetic material unit cell) that in the presence of non-negligible
losses one always should calculate the stored energy at the
microscopic level. 2) To generalize the previously reported
equivalent circuit representation for artificial magnetic media
\cite{Sergei_energy}. The generalized circuit model takes into
account losses in the matrix material. It is shown that this
generalization forces a modification to the widely accepted
permeability function used to macroscopically describe artificial
magnetic media. The generalization has a significant importance as
it is shown that in a practical situation matrix losses strongly
dominate over conductive losses. The proposed circuit representation
and permeability function are experimentally validated: We measure
the magnetic polarizability of a small magnetic material sample and
compare the results with those given by the proposed analytical
model and the previously used model. The results given by the
proposed model agree very well with the measured results, whereas
the old model leads to dramatic overestimation for the
polarizability. Using the generalized circuit model we derive an
expression for the field energy density in artificial magnetic
media. This expression is compared with the results obtained using
the ED-method in \cite{Boardman}. Reasons for the differences in the
results are discussed.

%\hfill

%It is verified using a specific example (magnetic material unit
%cell) that in the presence of non-negligible losses one always
%should calculate the stored energy at the microscopic level.

\section{Electrodynamic method vs.~Equivalent circuit method}

It is well explained in reference books
(e.g.~\cite{Landau,Vainstein}) that for the definition of field
energy density in a material having non-negligible losses one always
needs to know the material microstructure. First of all, the
reactive energy stored in any material sample is a quantity that can
be measured. When the material is \emph{lossless}, no information is
needed about the material microstructure for this measurement.
Indeed, we can measure the total power flux through the surface of
sample volume and, since there is no power loss inside, we can use
the Poynting theorem to determine the change in the stored energy.
This is the reason why the field energy density in a dispersive
media with negligible losses can be expressed through the frequency
derivative of macroscopic material parameters. In the circuit
theory, the same conclusion is true for circuits that contain only
reactive elements: It is possible to find the stored reactive energy in  the whole circuit
knowing only the input impedance of a two-pole \cite{Jackson}.

Simple reasoning reveals, that in the presence of non-negligible material losses the above described ``black
box'' representation and direct measurement are not applicable: Without knowing the material microstructure or
the circuit topology we do not know which portion of the input power is dissipated and which is stored in the
reactive elements. Thus, the energy stored in a lossy media cannot be \emph{uniquely} defined by only utilizing
the knowledge about the macroscopic behavior of the media \cite{Landau,Vainstein}.

\begin{figure}[b!]
\centering \epsfig{file=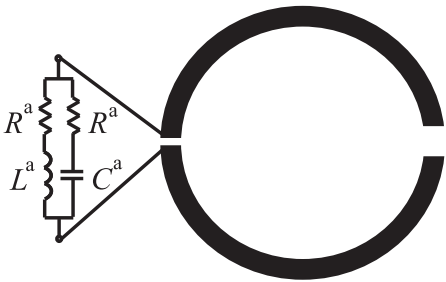, width=6.0cm} \caption{Split-ring resonator loaded with an infinitesimally
small lumped circuit.} \label{r}
\end{figure}

When defining the energy density in a certain material using the ED-method one first writes down the equation of
motion for charge density or for magnetization in the medium using the macroscopic media model \cite{Cui,Boardman}.
We stress that this equation is the \emph{macroscopic} equation of motion, containing the same physical information
as the macroscopic permittivity and permeability.
Further, complex
Poynting theorem is used to identify the mathematical form of the general macroscopic energy density
expressions. Having the form of these expressions in mind one searches for similar expressions in the equation
of motion and defines them as energy densities. The problem of the ED-method method is the fact that it only
utilizes the knowledge about the macroscopic behavior of the media, which, as explained above, in not
enough. The aforementioned difficulty is avoided in the EC-method \cite{Sergei_energy}. Based on the
known microscopic medium structure, one constructs the equivalent circuit for the unit cell of the medium.
Careful analysis is needed to make sure that the circuit physically corresponds to the analyzed unit cell. After
this check the stored reactive energy and the corresponding field energy density can be uniquely calculated
using the classical circuit theory.

Next, we illustrate the difference between the ED- and EC-methods using a specific case of split-ring
composites. Consider the split-ring resonator (SRR) shown in Fig.~\ref{r}. Following a example given in
\cite{Vainstein}, we load the SRR with an electrically infinitesimally small circuit consisting of lumped
elements. Let us assume that the additional inductance and capacitance are chosen to have values $L^{\rm a}=\tau
R^{\rm a}, C^{\rm a}=\tau/R^{\rm a}, \tau>0$. A simple check reveals that in this case the input impedance of
the load circuit is frequency independent and purely resistive: $Z_{\rm in}=R^{\rm a}$. When the SRR is
electrically small, it can be represented as a resonant contour and the total loss resistance reads $R_{\rm
tot}= R + R^{\rm a}$, where $R$ is the loss resistance due to the finite conductivity of the loop materials. Let
us further assume that the ring is made of silver and the value of $R^{\rm a}$ is chosen so that $R_{\rm tot}$
is the same as $R$ for an unloaded SRR made of copper. In this case the macroscopic descriptions of a medium
formed by the loaded SRRs made of silver with additional loads and unloaded SRRs made of copper are exactly the
same. Thus, the ED-method predicts the same value for the reactive energy stored in these two media. Inspection
of Fig.~\ref{r} clearly indicates, however, that this is not the case. There is additional energy stored in the
load inductance and capacitance, which is \emph{invisible} on the level of the macroscopic permeability
description. Proper definition of the stored energy must be done at the microscopic level, which is possible
with the equivalent circuit method.

\section{Equivalent circuit method: Brief revision of earlier results}

A commonly accepted permeability model as an effective medium description of dense (in terms of the wavelength)
arrays of split-ring resonators and other similar structures  reads
\e \mu(\omega) = \mu_0\mu_{\rm r}(\omega) =
\mu_0\bigg{(}1 + \frac{A\omega^2}{\omega_{\rm 0}^2 - \omega^2 + j\omega\Gamma}\bigg{)} \label{mu} \f (see
e.g.~\cite{Kostin, Pendry, Gorkunov, Maslovski}.)
Above, $A$ is the amplitude factor ($0<A<1$), $\omega_{\rm 0}$
is the undamped angular frequency of the zeroth pole pair (the resonant frequency of the array), and $\Gamma$ is
the loss factor. The model is obviously applicable only in the quasi-static regime since in the limit
$\omega\rightarrow\infty$ the permeability does not tend to $\mu_0$. At extremely high frequencies materials
cannot be polarized due to inertia of electrons, thus, a physically sound high frequency limit is $\mu_0$
\cite{Landau}. However, (\ref{mu}) gives correct results at low frequencies and
in the vicinity of the resonance. This is the typical frequency range of interest e.g.~when utilizing artificial
magneto-dielectric substrates in antenna miniaturization \cite{Mosallaei_disagree,Buell,Pekka}.
The other relevant restriction on the permeability function is the inequality \cite{Landau}
\e {\partial (\omega\mu)\over \partial\omega}>\mu_0, \l{1}\f
valid in the frequency regions with negligible losses. Physically the last restriction means that the
stored energy density in a passive linear lossless medium must be always larger than the energy density of the
same field in vacuum. Macroscopic model  (\ref{mu}) violates restriction \r{1} at high frequencies, which is
another manifestation of the quasi-static nature of the model.
In the vicinity
of the magnetic resonance the effective permittivity of a dense array of split-ring resonators is weakly
dispersive, and can be assumed to be constant.

\begin{figure}[t!]
\centering \epsfig{file=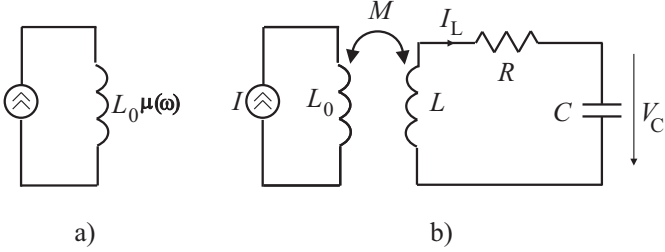, width=8.8cm} \caption{a) Magnetic material sample in the probe
magnetic field of a tightly wounded long solenoid. b) Equivalent circuit model, losses in the matrix material
are not taken into account.} \label{c_noloss}
\end{figure}

In \cite{Sergei_energy} the energy density in dispersive and lossy magnetic materials was introduced via a
thought experiment: A small (in terms of the wavelength or the decay length in the material) sample of a
magnetic material [described by (\ref{mu})] was positioned in the magnetic field created by a tightly wounded
long solenoid having inductance $L_0$, Fig.~\ref{c_noloss}a. The insertion changes the impedance of the solenoid
to \e Z(\omega) = j\omega{L_0}\mu_{\rm r}(\omega) = j\omega{L_0} + \frac{j\omega^3L_0A}{\omega_{\rm 0}^2 -
\omega^2 + j\omega\Gamma}. \label{Z1} \f The equivalent circuit with the same impedance was found to be that
shown in Fig.~\ref{c_noloss}b \cite{Sergei_energy} with the impedance seen by the source \e Z(\omega) =
j\omega{L_0} + \frac{j\omega^3M^2/L}{\frac{1}{LC} - \omega^2 + j\omega\frac{R}{L}}, \f which is the same as
(\ref{Z1}) if \e \frac{M^2}{LL_0}=A, \quad \frac{1}{LC} = \omega_{\rm 0}^2, \quad \frac{R}{L} = \Gamma. \f The
aforementioned equivalent circuit model is correct from the microscopic point of view since the modeled material
is a collection of capacitively loaded loops magnetically coupled to the incident magnetic field. An important
assumption in \cite{Sergei_energy} and in the present paper is that the current distribution is nearly uniform
over the loop. This means that the electric dipole moment created by the exciting field is negligible as
compared to the magnetic moment. The electromagnetic field energy density in the material was found to be
\cite{Sergei_energy} \e w_{\rm m} = \frac{\mu_0}{2}\bigg{(}1 + \frac{A\omega^2(\omega_{\rm 0}^2 +
\omega^2)}{(\omega_{\rm 0}^2 - \omega^2)^2 + \omega^2\Gamma^2}\bigg{)}|H|^2. \label{den1} \f

In \cite{Sergei_energy} only losses due to nonideally conducting metal of loops were taken into account, and
losses in the matrix material (substrate material on which metal loops are printed) were neglected. It will be
shown below that neglecting the matrix losses can lead to severe overestimation of the achievable permeability
values.

\section{Generalized equivalent circuit model and permeability function}

\begin{figure}[t!]
\centering \epsfig{file=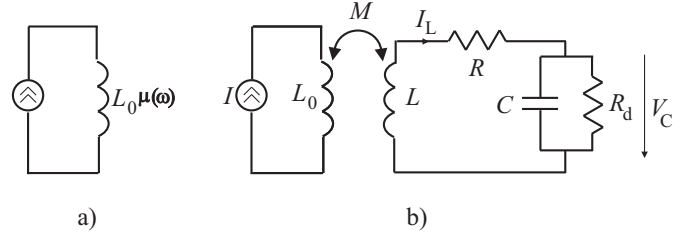, width=8.8cm} \caption{a) Magnetic material sample in the probe
magnetic field of a tightly wounded long solenoid. b) Equivalent circuit model taking into account losses in the
matrix material.} \label{c_loss}
\end{figure}

Losses in the matrix material (typically a lossy dielectric laminate) can be modeled by an additional resistor
in parallel with the capacitor. Indeed, if a capacitor is filled with a lossy dielectric material, the
admittance reads \e Y = j\omega{C}(\E' - j\E'') = j\omega{C}\E' + \omega{C}\E'', \label{lossr} \f where the
latter expression denotes a loss conductance. Thus, the microscopically correct equivalent circuit model is that
shown in Fig.~\ref{c_loss}b.
The impedance seen by the source can be readily solved: \e Z = j\omega{L_0} + \frac{j\omega^3M^2/L +
\omega^2M^2/(LCR_{\rm d})}{(1 + \frac{R}{R_{\rm d}})\frac{1}{LC} - \omega^2 + j\omega{(\frac{R}{L} +
\frac{1}{CR_{\rm d}})}}. \label{Zgen} \f The macroscopic permeability function corresponding to this model reads
\e \mu(\omega) = \mu_0\bigg{(}1 + \frac{\omega^2M^2/(LL_0) - j\omega{M^2}/(LL_0CR_{\rm d})}{(1 + \frac{R}{R_{\rm
d}})\frac{1}{LC} - \omega^2 + j\omega{(\frac{R}{L} + \frac{1}{CR_{\rm d}})}}\bigg{)}. \label{mu_gen} \f

Comparing (\ref{mu}) and (\ref{mu_gen}) we immediately notice that (\ref{mu}) is an insufficient macroscopic
model for the substrate if the losses in the host matrix are not negligible. A proper macroscopic model
correctly representing the composite from the microscopic point of view is
\e \mu(\omega) = \mu_0\mu_{\rm
r}(\omega) = \mu_0\bigg{(}1 + \frac{A\omega^2 - j\omega{B}}{\widetilde{\omega}_{\rm 0}^2 - \omega^2 +
j\omega(\Gamma + \Gamma_{\rm d})}\bigg{)}. \label{mu_G} \f Equation (\ref{mu_gen}) is the same as (\ref{mu_G})
if $$ \frac{M^2}{LL_0} = A, \quad \frac{M^2}{LL_0CR_{\rm d}} = B, \quad \bigg{(}1 + \frac{R}{R_{\rm
d}}\bigg{)}\omega_{\rm 0}^2 = \widetilde{\omega}_{\rm 0}^2,$$ \e \frac{R}{L} = \Gamma, \quad \frac{1}{CR_{\rm
d}} = \Gamma_{\rm d}. \label{param} \f Above we have denoted $\omega_{\rm 0}=1/(LC)$. The macroscopic
permeability function of different artificial magnetic materials can be conveniently estimated using
(\ref{mu_G}), as several results are known in the literature for the effective circuit parameter values for
different unit cells, e.g.~\cite{Pendry, Sauviac, Maslovski}.

For the use of (\ref{mu_G})  it is important to know the physical nature of the equivalent loss resistor $R_{\rm
d}$. If losses in the matrix material are due to finite conductivity of the dielectric material, the complex
permittivity reads \e \E = \E' - j\E'' = \E' - j\frac{\sigma}{\omega}, \label{perm1} \f where $\sigma$ is the
conductivity of the matrix material. Thus, we see from (\ref{lossr}) that the loss resistor is independent from
the frequency and can be interpreted as a ``true'' resistor. Moreover, in this case the permeability function is
that given by (\ref{mu_G}). However, depending on the nature of the dielectric material the loss mechanism can
be very different from (\ref{perm1}), and in other situations the macroscopic permeability function needs other
modifications. For example, let us assume that the permittivity obeys the Lorentzian type dispersion law \e \E =
\E'\bigg{(}1 + \frac{C}{\omega_0'^2 - \omega^2 + j\omega\Lambda}\bigg{)}, \label{perm2} \f where $\omega_0'^2$
is the angular frequency of the electric resonance, $C$ is the amplitude factor and $\Lambda$ is the loss
factor. Moreover, we assume that the material is utilized well below the electric resonance, thus,
$\omega\ll\omega_0'$. With this assumption the permittivity becomes \e \E \approx \E'(1 + C) -
j\omega{\E'C\Lambda}/\omega_0'^2. \label{perm3} \f We notice from (\ref{lossr}) that in this case the equivalent
loss resistor $R_{\rm d}$ becomes frequency dependent: \e R_{\rm d} \propto \frac{1}{\omega^2}, \label{Rmod} \f
and the permeability function takes the form \e \mu(\omega) = \mu_0\mu_{\rm r}(\omega) = \mu_0\bigg{(}1 +
\frac{A\omega^2 - j\omega^3{B'}}{\omega_0^2 - K\omega^2 + j\omega(\Gamma + \omega^2\Gamma_{\rm d}')}\bigg{)},
\label{mu_G_mod} \f where $K$ is a real-valued coefficient depending on the dielectric material. For other
dispersion characteristics of the matrix material the permeability function can have other forms.

\section{Experimental validation of the proposed circuit model and permeability function}

In this section we present experimental results that validate the generalized equivalent circuit model and the
corresponding macroscopic permeability function. The measurement campaign and the experimental results are described in
detail in \cite{Maslovski}. For the convenience of the reader we briefly revise the main steps of the
measurement procedure.

\begin{figure}[b!] \centering \epsfig{file=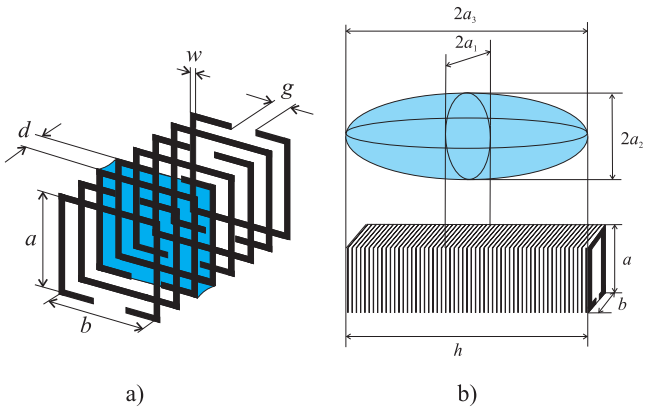, width=8.5cm} \caption{a) A schematic
illustration of the metasolenoid. b) A finite-size metasolenoid approximated as a magnetic ellipsoid.}
\label{metasolenoid}
\end{figure}

The measured artificial magnetic particle, metasolenoid, is schematically presented in Fig.~\ref{metasolenoid}a.
In \cite{Maslovski} the effective permeability of a medium densely filled with infinitely long metasolenoids was
derived in the form \e \mu_{\rm eff} = 1 - V_{\rm r}\frac{j\omega\mu_0 S}{Z_{\rm tot} d}, \label{mu_eff} \f
where $V_{\rm r}$ is the volume filling ratio, $S=a\times b$ is the cross section are of the ring, $d$ is the
distance between the rings, and the total effective impedance was presented in the form \e Z_{\rm tot} =
j\omega{L_{\rm eff}} + \frac{1}{j\omega C_{\rm eff}} + R_{\rm eff}. \label{Z} \f For the experimental validation
a finite-size metasolenoid was approximated as an ellipsoid cut off from a magnetic media described by
(\ref{mu_eff}), Fig.~\ref{metasolenoid}b. The magnetic polarizability of the ellipsoid $\alpha_{\rm mm}$ was
analytically calculated using the classical mixing theory \cite{Ari}. The scattering amplitude of an electrically
small material sample was measured using a standard parallel plate waveguide, and the theory of waveguide
excitation was used to calculate the magnetic polarizability from the measured results. It is worth noting that
the magnetic polarizability is a function of the magnetic permeability due to the mixing process. On the other
hand, permeability is defined using the equivalent circuit, Fig.~\ref{c_loss}b. Thus, the magnetic polarizablity
of the measured sample contains all the relevant data for validating both the proposed equivalent circuit and
the permeability function.

\begin{figure}[b!] \centering \epsfig{file=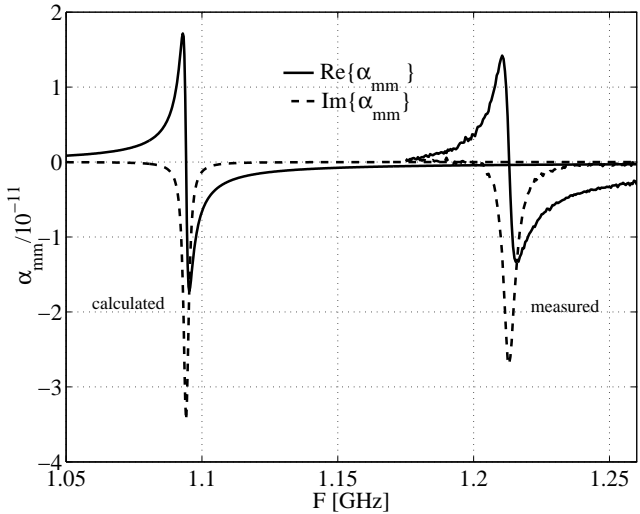, width=8.5cm} \caption{The analytically calculated
(proposed model) and measured magnetic polarizability.} \label{meas}
\end{figure}

Though it is not explicitly mentioned in \cite{Maslovski}, substrate losses were taken into account when
analytically calculating the magnetic polarizability of the sample. The authors used (\ref{Z}) to define the
total impedance of the metasolenoid unit cell, however, complex permittivity was used when calculating the
effective capacitance. Thus, the equivalent circuit of the unit cell used to analyze the measured sample is the
proposed circuit shown in Fig.~\ref{c_loss}b. It can easily be verified using the data presented in
\cite{Maslovski} that the following expression for the effective impedance (derived using the circuit in
Fig.~\ref{c_loss}b) exactly repeats the analytical estimation for the magnetic polarizability of the sample: \e
Z_{\rm tot}' = j\omega L_{\rm eff} + R_{\rm eff} + \frac{R_{\rm d}}{1 + j\omega C_{\rm eff}R_{\rm d}},
\label{Zmod} \f where $R_{\rm d}=\E'/(\E''\omega C_{\rm eff})$. The analytically calculated [$Z_{\rm tot}$ given
by (\ref{Zmod}) is used in (\ref{mu_eff})] and measured magnetic polarizabilities are shown in Fig.~\ref{meas}.
The measured and calculated key parameters are gathered in Table~\ref{t_meas}. The polarizability and permeability
values in Table~\ref{t_meas} are the maximum values.

The measured results agree rather closely with the analytical calculations when the proposed model is
used. The slight difference in the resonant frequencies, and the slightly lowered polarizability values in the
measurement case are most likely caused by limitations in the accuracy of the manufacturing process: The
implemented separation between the rings is probably slightly larger than the design value. This lowers the
effective capacitance and is seen as a weakened magnetic response and the higher resonant frequency. Moreover, the
measured Im$\{\alpha_{\rm mm}\}$ clearly indicates that the effect of the lossy glue used to stack the rings is
underestimated in the analytical calculations. In the analytical calculations $\tan\delta=0.002$ was used for
the total loss tangent \cite{Maslovski}. A loss tangent value $\tan\delta=0.0025$ would accurately produce the
measured polarizability values, and the bandwidths (defined from the Im$\{\alpha_{\rm mm}\}$ curve) would
visually coincide.

If the matrix losses are neglected in the circuit model [$R_{\rm d}\rightarrow\infty$ in (\ref{Zmod})], the
analytical calculations lead to dramatically overestimated polarizability and permeability values. It is
therefore evident that the proposed generalization of the circuit model and the permeability function have a
significant practical importance. Though the model has been validated using a specific example, we can conclude
that matrix losses can strongly dominate over conductive losses in structures where the unit cells are
closely spaced. This is physically well understandable since in this case most of the flux is forced inside the
substrate.

\begin{table}[t!]
\centering \caption{Analytically calculated and measured properties of the metasolenoid sample.} \label{t_meas}
\begin{tabular}{|c|c|c|c|c|}
\hline & $F_{\rm res}$  & Re$\{\alpha_{\rm mm}\}$ & $|\alpha_{\rm mm}|$ & Re$\{\mu_{\rm eff}\}$ \\
 & GHz & H$\cdot$m$^2$ & H$\cdot$m$^2$ & \\
\hline
Analytical$^\dag$ & 1.09 & 1.7$\times10^{-11}$ & 3.3$\times10^{-11}$ & 230 \\
Measured & 1.21 & 1.4$\times10^{-11}$ & 2.7$\times10^{-11}$ & --- \\
Analytical$^\ddag$ & 1.09 & 19.9$\times10^{-11}$ & 60.1$\times10^{-11}$ & 4000 \\\hline
\end{tabular}
\smallskip
$^\dag${\small proposed model,} $^\ddag${\small old model}
\end{table}

\section{Electromagnetic field energy density}

Following the approach introduced in \cite{Sergei_energy} we will next generalize the expression for the energy
density in artificial magnetics using the experimentally validated circuit model. In the time-harmonic regime
the total stored energy reads (notations are clear from Fig.~\ref{c_loss}b) $$ W = \frac{1}{2}(L_0|I|^2 +
L|I_{\rm L}|^2 + C|V_{\rm C}|^2)$$ \e = \frac{1}{2}\bigg{[}L_0|I|^2 + |I_{\rm L}|^2\bigg{(}L +
\frac{1}{C(\omega^2 + \frac{1}{C^2R_{\rm d}^2})}\bigg{)}\bigg{]}, \f \e |I_{\rm L}|^2 =
\frac{\omega^2\frac{M^2}{L^2}\bigg{(}\omega^2 + \frac{1}{C^2R_{\rm d}^2}\bigg{)}}{[(1 + \frac{R}{R_{\rm
d}})\frac{1}{LC} - \omega^2]^2 + \omega^2{(\frac{R}{L} + \frac{1}{CR_{\rm d}})^2}}|I|^2. \label{ILabs} \f Using
the notations in (\ref{param}) the stored energy can be written as
\e W = \frac{1}{2}L_0|I|^2\bigg{(}1 +
\frac{A\omega^2(\omega_{\rm 0}^2 + \omega^2 + \Gamma_{\rm d}^2)}{(\widetilde{\omega}_{\rm 0}^2 - \omega^2)^2 +
\omega^2(\Gamma + \Gamma_{\rm d})^2}\bigg{)}. \label{W} \f The inductance per unit length of a tightly wound
long solenoid is $L_0=\mu_0n^2S$, where $n$ is the number of turns per unit length and $S$ is the cross section
area. The relation between the current $I$ and magnetic field $H$ inside the solenoid is $I=H/n$. Thus, the
stored energy in one unit-length section of the solenoid reads \e W = w_{\rm m}S =
\frac{1}{2}\mu_0n^2S\frac{|H|^2}{n^2}\bigg{(}1 + \frac{A\omega^2(\omega_{\rm 0}^2 + \omega^2 + \Gamma_{\rm
d}^2)}{(\widetilde{\omega}_{\rm 0}^2 - \omega^2)^2 + \omega^2(\Gamma + \Gamma_{\rm d})^2}\bigg{)},
\label{Wfinal} \f from which we identify the expression for the electromagnetic field energy density in the
artificial material sample: \e w_{\rm m} = \frac{\mu_0}{2}\bigg{(}1 + \frac{A\omega^2(\omega_{\rm 0}^2 +
\omega^2 + \Gamma_{\rm d}^2)}{(\widetilde{\omega}_{\rm 0}^2 - \omega^2)^2 +\omega^2(\Gamma + \Gamma_{\rm
d})^2}\bigg{)}|H|^2. \label{wmfinal} \f We immediately note that if there is no loss in the matrix material
($R_{\rm d}\rightarrow\infty$ and $\Gamma_{\rm d}\rightarrow0$), then $\widetilde{\omega}_{\rm
p}^2\rightarrow\omega_{\rm p}$ and (\ref{wmfinal}) reduces to (\ref{den1}).

\subsection{Comparison with the results obtained using the ED-method}

The above derived result differs from the result found in \cite{Boardman} using the ED-method: \e w_{\rm m} =
\frac{\mu_0}{2}\bigg{(}1 + \frac{A\omega^2[\omega_{\rm 0}^2(3\omega_0^2 - \omega^2) +
\omega^2\Gamma^2]}{\omega_{\rm 0}^2[(\omega_0^2 - \omega^2)^2 +\omega^2\Gamma^2]}\bigg{)}|H|^2.
\label{wmfinal_comp} \f The procedure and the underlying assumptions to obtain (\ref{wmfinal_comp}) have been
briefly revised in Section II. The classical expression for the magnetic field energy density in a media where
absorption due to losses can be neglected reads \cite{Landau,Vainstein} \e w_{\rm m} =
\frac{\mu_0}{2}\frac{\partial(\omega\mu_{\rm r}(\omega))}{\partial\omega}|H|^2. \label{classic} \f It is seen
that in the presence of negligible losses [$\Gamma\rightarrow0$ in (\ref{mu})] the energy density result given
by (\ref{wmfinal_comp}) is the same as the result predicted by the classical expression (\ref{classic}).
However, (\ref{wmfinal}) predicts a different result. Authors of \cite{Boardman} use this fact to state that the
result obtained using the ED-method is more internally consistent than the result obtained using the EC-method.

The EC-method is known to give a perfectly internally consistent result for the energy density in dielectrics
obeying the general Lorentzian type dispersion law \cite{Sergei_energy}. The general Lorentz model is a strictly
causal model. This is, however, not the case with the modified Lorentz model (\ref{mu}). As is speculated
already in \cite{Sergei_energy}, the reason for the difference in the results obtained using (\ref{wmfinal}) and
(\ref{classic}) in the small-loss limit is related to the physical limitations of the quasi-static permeability
model (\ref{mu}). Thus, when (\ref{mu}) is used as the macroscopic media description, (\ref{wmfinal}) should be
used also in the presence of vanishingly small losses. The ED-method, though being internally consistent with
the classical expression, predicts unphysical behavior at high frequencies: At high frequencies the energy
density given by the ED-method is smaller than the energy density in vacuum (when there are no losses, this
unphysical behavior takes place at frequencies $\omega>\sqrt{3}\omega_{\rm 0}$, where restriction \r{1}
is violated), Fig.~1a and 1b in
\cite{Boardman}. This behavior is avoided with the result obtained using the EC-method, since that approach is
based on the microscopic description of the medium, which is always in harmony with the causality principle.

\begin{figure}[t!]
\centering \epsfig{file=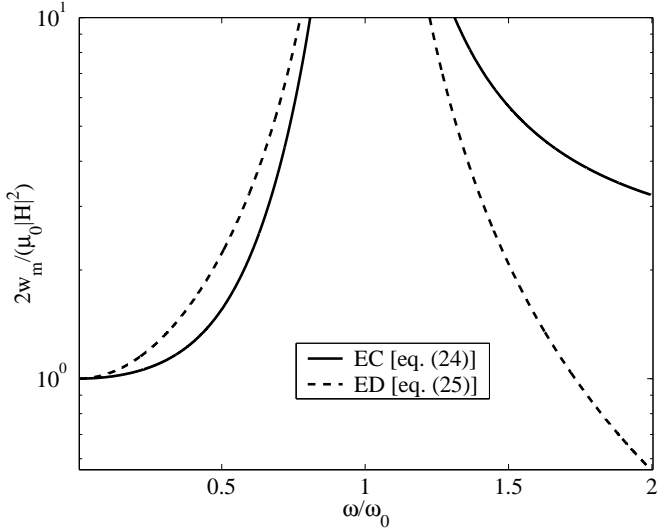, width=8.8cm} \caption{Magnetic field energy density given by different
expressions. $\omega_0$ is the resonant frequency of the metasolenoid medium.} \label{wmp}
\end{figure}

Fig.~\ref{wmp} depicts the normalized magnetic field energy density in a medium formed by the metasolenoids
introduced in the previous section. The amplitude factor $A=1$, and the loss factors have been estimated using
(\ref{param}) and the data presented in \cite{Maslovski}. In this particular example the energy densities given
by (\ref{wmfinal}) and (\ref{den1}) give practically the same result over the whole studied frequency range (the
result given by (\ref{den1}) is not plotted in Fig.~\ref{wmp} since the result visually coincides with the
EC-result). This is due to the fact that large values of $\omega$ and $\omega_{\rm 0}$ mask the effect of
$\Gamma$ and $\Gamma_{\rm d}$ in (\ref{wmfinal}) and (\ref{den1}). The results given by the ED-method and the
classical expression (\ref{classic}) also visually coincide. However, as was mentioned above, the energy density
expression given by the ED-method predicts the same nonphysical behavior as the classical expression: The field
energy density is smaller than the energy density in vacuum at frequencies $\omega>\sqrt{3}\omega_{\rm 0}$.

\section*{Conclusion}

In this paper we have explained differences between recent approaches used to derive field
energy density expressions for artificial magnetic media. The equivalent circuit model of split-ring resonators
and other similar structures has been generalized to take account losses in the dielectric matrix material. It
has been shown that a modification is needed to the macroscopic permeability function commonly used to model
these materials in the quasi-static regime. Moreover, depending on the nature of the dominating loss mechanism
in the matrix material, the permeability function has different forms. The proposed circuit model and the modified
permeability function have been experimentally validated, and it has been shown that in a practical situation
matrix losses can dramatically dominate over conductive losses. Using the validated circuit model we have
derived an expression for the electromagnetic field energy density in artificial magnetic media. This expression
is valid also when losses in the material cannot be neglected and when the medium is strongly dispersive. The
results have been compared to the recently reported results.

\section*{Acknowledgements}

This work has been done within the frame of the European Network
of Excellence {\itshape Metamorphose}.  We would like to
acknowledge financial support of the Academy of Finland and TEKES
through the Center-of-Excellence program and thank Professor
Constantin Simovski for useful discussions.

\end{document}